\documentclass[12pt]{article}
\usepackage{amsmath,amssymb,exscale}
\usepackage{graphicx}
\usepackage{epsfig}
\usepackage{multicol}
\usepackage{color}
\usepackage{mathrsfs }
\usepackage{blindtext}
 \usepackage{fancyhdr}
\usepackage{hyperref}
\usepackage{cite}

\numberwithin{equation}{section}

\usepackage{xcolor}
\usepackage{sectsty}

\usepackage{mdframed}
\usepackage{titletoc}

\definecolor{secnum}{RGB}{13,151,225}
\definecolor{ptcbackground}{RGB}{212,237,252}
\definecolor{ptctitle}{RGB}{0,177,235}

\titlecontents{lsection}
  [5.8em]{\sffamily}
  {\color{secnum}\contentslabel{2.3em}\normalcolor}{}
  {\titlerule*[1000pc]{.}\contentspage\\\hspace*{-5.8em}\vspace*{5pt}%
    \color{white}\rule{\dimexpr\textwidth-15.5pt\relax}{1pt}}

\usepackage{hyperref}
\hypersetup{colorlinks,bookmarksopen,bookmarksnumbered,citecolor=rossos,
linkcolor=redy,pdfstartview=FitH,urlcolor=green-go}
\usepackage{slashed}

\definecolor{blus}{cmyk}{1,1,0,0.1}
\definecolor{verdes}{cmyk}{0.99,0,0.59,0.65}
\definecolor{rossos}{cmyk}{0,1,1,0.55}
\definecolor{redy}{cmyk}{0,1,1,0.7}
\definecolor{greeny}{cmyk}{0.99,0,0.59,0.98}
\definecolor{green-go}{cmyk}{0.79,0,0.59,0.5}

\usepackage{titlesec}

\def\hhref#1{\href{http://arxiv.org/abs/#1}{arXiv:#1}} 

\newcommand{\tmtextbf}[1]{{\bfseries{#1}}}
\newcommand{\tmtextrm}[1]{{\rmfamily{#1}}}

\newcommand{\gs}{f_0}
\newcommand{\gt}{f_2}

\def\be{\begin{equation}}
\def\ee{\end{equation}}
\def\ba{\begin{array} }
\def\bac{\begin{array} {c}}
\def\bacc{\begin{array} {cc}}
\def\baccc{\begin{array} {ccc}}
\def\bacccc{\begin{array} {cccc}}
\def\ea{\end{array}}
\def\bea{\begin{eqnarray}}
\def\eea{\end{eqnarray}}

\definecolor{red}{rgb}{1,0,0}

\def\psl{\hbox{\hbox{${p}$}}\kern-1.9mm{\hbox{${/}$}}}
\def\dsl{\hbox{\hbox{${\partial}$}}\kern-2.2mm{\hbox{${/}$}}}
\def\Dsl{\hbox{\hbox{${D}$}}\kern-2.6mm{\hbox{${/}$}}}
\newcommand{\bp}{\bar M_{\rm Pl}}

\def\Lag{\mathscr{L}}

\newcommand{\gappeq}{{\rlap{{\raise}.5ex\text{\ensuremath{>}}}{{\lower}.5ex\text{\ensuremath{\sim}}}}}
\newcommand{\lappeq}{{\rlap{{\raise}.5ex\text{\ensuremath{<}}}{{\lower}.5ex\text{\ensuremath{\sim}}}}}

\newcommand{\I}{\tmtextrm{1{\kern}-.24em l}}

\begin{document}
\topmargin -1.0cm
\oddsidemargin -0.5cm
\evensidemargin -0.5cm

\vspace{-1cm}

 \begin{center}CERN-TH-2016-176\end{center}
 
{\vspace{0.3cm}}
\begin{center}

 {\huge  \tmtextbf{ 
{\color{blus} Solving the Standard Model Problems in Softened Gravity  }}} {\vspace{.5cm}}\\

\vspace{0.7cm}

{\large  {\bf Alberto Salvio
}
\vspace{-.4cm}

{\it }\

{\em  \normalsize 

CERN, Theory Division, CH-1211 Geneva 23, Switzerland\\

\vspace{0.2cm}

 \vspace{0.5cm}
}

}
\vspace{1.9cm}
 \end{center}
\noindent --------------------------------------------------------------------------------------------------------------------------------

\vspace{-0.5cm}

\begin{center}
{\bf  Abstract}
\end{center}

\noindent {\normalsize    The Higgs naturalness problem is solved if the growth of  Einstein's gravitational interaction is softened at an energy  $ \lesssim 10^{11}\,$GeV  ({\it softened gravity}). We work here within an explicit realization  where the Einstein-Hilbert Lagrangian is extended to include terms quadratic in the curvature and a non-minimal coupling with the Higgs. We show that this solution is preserved by adding three right-handed neutrinos with masses below the electroweak scale, accounting for neutrino oscillations, dark matter and the baryon asymmetry. The smallness of  the right-handed neutrino masses (compared to the Planck scale) and the QCD $\theta$-term are also shown to be natural. We prove that a possible gravitational source of CP violation cannot spoil the model, thanks to the presence of right-handed neutrinos. 
Inflation is approximately described by the Starobinsky model in this context, and can occur even if we live in a metastable vacuum.
}


\vspace{.2cm}

\noindent --------------------------------------------------------------------------------------------------------------------------------

\vspace{-0.3cm}


\newpage

{\color{redy}
\tableofcontents
}

%

\newpage

\section{Introduction and summary} \label{introduction and summary}

 The  hierarchy problem consists in finding an extension of the Standard Model (SM) where the Higgs mass $M_h$ is natural: quantum corrections to $M_h$ are small compared to its observed value. A challenge is to achieve this in the presence of gravity.  Softened gravity is a scenario in which the growth of Einstein's gravitational interaction stops at a scale no larger than $10^{11}\,$GeV \cite{Giudice:2014tma}. In such a situation  the gravitational quantum corrections to $M_h$ are not too large solving the  hierarchy problem. An important question is whether this scenario can be made realistic and can address the shortcomings of the SM: non-zero neutrino masses, dark matter (DM), baryon asymmetry of the universe (BAU),  inflation as well as an explication for the smallness of the QCD $\theta$-term.
 
 Here we show that this can be achieved by simply including three right-handed neutrinos with Majorana masses $M$ below the EW scale. Right-handed neutrinos can account for the observed neutrino oscillations, DM and BAU. We consider a concrete 
implementation of the softened gravity idea where the Einstein-Hilbert Lagrangian is extended to include all terms quadratic in the curvature as well as a non-minimal coupling $\xi$ between the Higgs and gravity. We address the question of whether this theory might be a candidate UV completion of Einstein gravity.

 The same radiatively stable values of the parameters that lead to a natural Higgs mass (found in \cite{Salvio:2014soa}), also preserve the smallness of $M$ and $\theta$. The concept of naturalness used here is the one based on finite quantities (after renormalization), where unphysical power-law divergences with respect to the momentum cutoff are disregarded \cite{Vissani:1997ys, Farina:2013mla,deGouvea:2014xba,Clarke:2016jzm}. 
 
 We also show that a possible gravitational breaking of CP (which could be due to $\theta$) produces no visible effects in the observable quantities, thanks to the presence of the right-handed neutrinos. Inflation is mainly due to the effective Starobinsky scalar $z$ \cite{Starobinsky} (which automatically emerge from the terms quadratic in the curvature) and the Higgs gives very small contributions even in the natural parameter space. 
 
 We find it remarkable that all the above-mentioned problems can be solved in such a simple extension of the SM. 
 
 This paper is organized as follows. In the next section we describe the theory of softened gravity that will be considered in detail in this paper. In section \ref{The spectrum} we discuss its spectrum. The following section \ref{Quantum aspects} is dedicated to the study of some quantum aspects. Section \ref{nat} shows the naturalness of the Higgs mass, the QCD $\theta$ angle and right-handed neutrino masses and Yukawa couplings in this context. In the same section we also discuss the connection between the possible gravitational violation of CP and the neutrino sector. Section \ref{inf} presents a detailed analysis of inflation and finally we provide our conclusions and outlook in section \ref{Conclusions and outlook}. We provide technical material in two appendices.


\section{The theory}\label{The model}
 
   The full Lagrangian (density) is given by 
 \be \mathscr{L} =\sqrt{-g}\left( \mathscr{L}_{\rm gravity}+\mathscr{L}_{\rm SM}+\mathscr{L}_N\right). 
\label{full-lagrangian} \ee 
Here  $\mathscr{L}_{\rm gravity}$ represents the pure gravitational Lagrangian plus the possible non-minimal coupling between the Higgs  and gravity, which, modulo total derivatives \cite{Stelle:1977ry}, is
\be \hspace{3.9cm}\mathscr{L}_{\rm gravity}
= \frac{R^2}{6\gs^2} + \frac{\frac13 R^2 -  R_{\mu\nu}^2}{\gt^2} -\left(\frac{\bp^2}{2}+\xi |H|^2\right)R - \Lambda,  \phantom{+ \left( \frac{1}{f_{\rm  GB}^2} -\frac{1}{2f_2^2}\right)\left(R^2 +  R_{\mu\nu\rho\sigma}^2-4R_{\mu\nu}^2\right)} \hspace{-3.9cm} \label{gravity-Lag}
  \ee
  where $f_2$ and $f_0$ play the role of gravitational couplings, $\bp$ is the reduced Planck mass and $\Lambda$ is the cosmological constant. $\mathscr{L}_{\rm SM}$ represents the usual SM Lagrangian, minimally coupled to gravity.
$\mathscr{L} _{N}$ is the term that depends on the right-handed neutrinos $N_i$ (i=1,2,3):
\be \mathscr{L} _{N}= i\overline{N}_i \dsl N_i+ \left(\frac12 N_i M_{ij}N_j +  Y_{ij} L_iH N_j + {\rm h.  c.}\right) ,\ee
where $M_{ij}$ and $Y_{ij}$  are the elements of  the Majorana mass matrix $M$
and the neutrino Yukawa coupling matrix $Y$, respectively. 

Note that the new gravitational terms associated with the couplings $f_0$, $f_2$ and $\xi$ are necessary for the renormalization of the theory: even if we do not introduce them in the classical theory radiative corrections generate them.

\section{The spectrum}\label{The spectrum}

 We now turn to the spectrum.  As usual there is a massless spin-2 graviton.  Also, the term $(\frac13 R^2 -  R_{\mu\nu}^2)/f_2^2$ in (\ref{gravity-Lag}) corresponds to a ghost: a field with an unusual minus sign in front of its kinetic term. In the presence case this field has spin-2 and  mass $M_2 \equiv f_2 \bp/\sqrt{2}$ \cite{Stelle:1976gc}. In the next section we will discuss a possible sensible way of interpreting such a field.

 The term $R^2/(6f_0^2)$ leads instead to the scalar $z$. To see this one  can write the scalar-tensor part of the Lagrangian, $\Lag_{\rm st}$, in the $E$instein frame \cite{Salvio:2014soa,Kannike:2015apa,Salvio:2015kka}
\be \label{eq:Einstein}
\Lag_{\rm st} =\sqrt{- g_E} \bigg[  \frac{\frac13 R_E^2 -  R_{E\mu\nu}^2}{f_2^2}
 - \frac{\bar M_{\rm Pl}'^2}{2} R_E +  \Lag_\phi - V_E  \bigg],
\ee
where everything is computed with the new metric
\be g^E_{\mu\nu}\equiv  g_{\mu\nu}\times  \frac{z^2}{6\bp'^2}, \ee
$\bar M_{\rm Pl}'^2 \equiv \bp^2+2\xi v^2$ (note that $v\simeq 174\,$GeV and in practice one can take $\bar M_{\rm Pl}'^2 = \bar M_{\rm Pl}^2$) and 
\be \Lag_\phi \equiv
  \frac{6\bp'^2}{z^2}\left(|D_\mu H|^2 + 
 \frac{(\partial_\mu z)^2}{2}\right)  ,\qquad
V_E \equiv \frac{36\bp'^4}{z^4}
\bigg[{V(H)}+   \frac{3f_0^2}{8} \bigg(\frac{z^2}{6} - 2 \xi |H|^2-\bp^2\bigg)^2\bigg]. \label{scalar-EF}  \ee
The potential $V$ is the SM one: $V(H) = \lambda (|H|^2-v^2)^2 + \Lambda$.

The  minimum of $V_E$ occurs when the Higgs is at the electroweak (EW) scale, $v$,  and 
$z\approx \langle z\rangle \equiv \sqrt{6\bp'^2}$ (here we neglect tiny corrections due to  $\Lambda/\bp\neq 0$). Notice that at $z=\langle z\rangle$ the kinetic terms of the scalars are canonically normalized, therefore the squared mass matrix for scalars, $\mathscr{M}_0^2$,  is the Hessian matrix of $V_E$ computed at this point of minimum. This procedure leads to 
\be  \mathscr{M}_0^2 \approx \left(\begin{array}{cc}
M_0'^2 &   -\epsilon M_0' m\\  
-\epsilon M_0' m  & (1+\epsilon^2)m^2
\end{array}\right),\nonumber\ee 
where    $M_0'^2\equiv f_0^2 \bp'^2/2$, $m\equiv 2\sqrt{\lambda} v$
and $  \epsilon\equiv \sqrt{\frac{6}{4\lambda}} f_0 \xi$
(note that   $\lambda$ is required to be positive by the stability of the potential $V$). As usual, when $\lambda <0$ we have a tachyonic instability\footnote{Note, however, that the possible metastability of the EW vacuum (which corresponds to $\lambda<0$ at very high field values)  does not rule out this model because the corresponding life-time  exceeds the age of the universe \cite{Buttazzo:2013uya,Salvio:2016mvj}.}. For $M_0'\gg m$, which we expect because  $M_0' \propto\bp$, 
the mixing becomes small and the scalar masses are approximately $M_0'$ and $m$. We will see in section \ref{Higgs-Starobinsky system} that this approximation is very accurate.

To analyse  the neutrino sector we  take (thanks to the complex Autonne-Takagi factorization)
 $M$ real and diagonal without loss of generality: 
$M=\mbox{diag}(M_1, M_2, M_3),$ where the $M_i$ are real mass parameters. The neutrinos acquire a Dirac mass matrix  $m_D = v Y,$
  which can  be parameterized as 
$ m_D =\left(\begin{array}{ccc}\hspace{-0.1cm}m_{D1}\,, & \hspace{-0.2cm}m_{D2}\, ,  & \hspace{-0.2cm} m_{D3}\hspace{-0.1cm}
\end{array}\right), $
where $m_{Di}$  are column vectors.   Integrating out the heavy neutrinos $N_i$, one then obtains  the usual see-saw formula for the light neutrino Majorana mass matrix  
\be m_\nu =  \frac{m_{D1} m_{D1}^T}{M_1} + \frac{m_{D2} m_{D2}^T}{M_2} + \frac{m_{D3} m_{D3}^T}{M_3} . \label{see-saw} \ee 
  By means of a unitary (Autonne-Takagi) redefinition of the left-handed SM neutrinos we can diagonalize $m_\nu$ to obtain the mass eigenvalues $m_1, m_2$ and $m_3$ (the left-handed neutrino Majorana masses). The experimental constraints on neutrino masses and oscillations (see  \cite{Capozzi:2016rtj}  for recent determinations) can be satisfied by choosing appropriately the unitary matrix $U_\nu$ that implements such transformation (the PMNS matrix) and the $m_i$ . 
  
  Here the $N_i$ are also responsible for DM \cite{Dodelson:1993je,Asaka:2005pn,Canetti:2012kh} and BAU \cite{Akhmedov:1998qx,Asaka:2005pn,Canetti:2012kh}. For example we find that all bounds to account for neutrino masses and oscillations (within 1$\sigma$)  \cite{Capozzi:2016rtj}, for DM and BAU   \cite{Canetti:2012kh} can be satisfied  for
  \be M_1 \sim {\rm keV}, \quad M_{2,3} \sim {\rm GeV}, \quad |Y_{ij}|< 10^{-7}. \label{RHvalue} \ee 
This low-scale right-handed neutrinos can be searched in the laboratory or through astrophysical observations (see for example  \cite{Canetti:2012kh,Adhikari:2016bei} and reference therein). This proposal can therefore be tested.

 \section{Quantum aspects}\label{Quantum aspects}

The theory with Lagrangian (\ref{full-lagrangian}) by itself does not eliminate the Landau poles of the SM, which, however, occur many orders of magnitude above the Planck scale where no experiments or observations can be made. We will therefore avoid this problem by assuming that there is a minimal length  much larger than the Landau pole scales, but still much shorter than the Planck length.

 The Lagrangian in (\ref{full-lagrangian}) defines a renormalizable theory of gravity \cite{Stelle:1976gc} (actually of all interactions).   The price to pay   is the presence of a ghost (which we have seen in the previous section). Such a field emerges because of the presence of 4 time-derivatives in the Lagrangian \cite{Ostro}. The spectrum of such a theory, however,  becomes bounded from below at the {\it quantum level} if negative norm states are introduced (see e.g. \cite{Salvio:2015gsi}  and references therein). In particular the quanta of the ghost field have to have negative norm, while all remaining quanta  have  positive norm. Recently, Ref. \cite{Salvio:2015gsi} showed (assuming a single 4-derivative degree of freedom) that such a  quantization can be obtained with normalizable wave-functions and a well-defined Euclidean path-integral. 
 
 The remaining problem of having a sensible probabilistic interpretation of negative-norm states could be bypassed by the Lee-Wick idea \cite{Lee:1969fy}, which assumes  all stable states in the theory to have positive norm. Indeed, theories of this sort are sensible as long as we only look at the energy spectrum and transition probabilities  between asymptotic (stable) states ($S$-matrix elements). Ref. \cite{Antoniadis} argued that the assumption of Lee and Wick is satisfied in  the theory of gravity above. One of the purposes of this section is to  {\it explicitly prove} that the ghost is unstable in this theory.
 
 \subsection{Ghost decay}
 
 Given that the ghost at hand has spin-2, a direct calculation of its decay involves the complications of the corresponding Lorentz indices. For this reason one would like to use the optical theorem and compute equivalently the imaginary part of the ghost propagator. However, the optical theorem is derived in theories with positive norms only and requires a generalization here, given that the ghost state has negative norm. 
 
 To obtain such generalization consider the time evolution operator $U$, which is defined as usual as the linear operator that transforms the state at the initial time into the state at the generic time. The usual procedure is to define the operator $T$ by 
 \be U\equiv 1+iT\ee 
 By using the unitarity condition (which is fulfilled even in the presence of negative norms \cite{Salvio:2015gsi}) we obtain
\be i(T^\dagger-T)=T^\dagger T.\ee
If we now take the matrix element between an initial state $|i\rangle$ and a final state $|f\rangle$ we find 
\be i(T^\dagger_{fi}-T_{fi})= (T^\dagger T)_{fi} \label{optical1}\ee
where $T_{fi}\equiv \langle f |T| i\rangle$, $T^\dagger_{fi}\equiv \langle f |T^\dagger| i\rangle$ and $(T^\dagger T)_{fi}\equiv  \langle f |T^\dagger T| i\rangle$. In theories with positive norms only the completeness relation is $1=\sum_n|n\rangle\langle n|$, where $\{|n\rangle\}$ is an orthonormal basis, $\langle n'|n\rangle=\delta_{n'n}$. In the presence of both negative and positive norms, however, the scalar product between two generic states $|\alpha\rangle$ and $|\beta\rangle$  can be written as 
 \be \langle \beta|\alpha\rangle  = (\beta, \eta \alpha ), \label{2scalarProducts}\ee
 where $(. \, , .)$ is a positively defined scalar product and $\eta$ is assumed to be a  diagonalizable operator with eigenvalues $+1$ and $-1$ \cite{Pauli1943}, so $\eta^2=1$. The completeness relation now reads $\eta=\sum_n|n\rangle \langle n|$, so
\be 1= \eta^2=\sum_n|n\rangle \langle n| \eta=\sum_n|n\rangle \langle n| \eta_n,\ee
where we have constructed the basis $\{|n\rangle\}$ with the eigenvectors of $\eta$. By inserting this into Eq. (\ref{optical1}) we get
\be i(T^\dagger_{fi}-T_{fi})=\sum_n \eta_n T_{nf}^* T_{ni},\ee
which is the generalization of the optical theorem we were looking for. 

To apply this formula to compute a decay we  set the initial and final states  equal to each other ($|i\rangle =|f\rangle = |\alpha\rangle$) and so 
\be 2\, {\rm Im}\left(T_{\alpha\alpha}\right)=\sum_n\eta_n|T_{n\alpha}|^2.\label{optical2}\ee
Notice that $|T_{n\alpha}|^2=|U_{n\alpha}-\langle n|\alpha \rangle|^2$, where $U_{n\alpha}\equiv \langle n|U|\alpha\rangle$. 
We are interested here in the decay of a negative norm particle; we therefore focus on states $|\alpha\rangle$ that are normalizable and set  $\langle \alpha|\alpha\rangle= \pm 1$. We would like to apply Eq. (\ref{optical2}) within perturbation theory at first order, so we can take   $T_{\alpha\alpha}=0$
and $\eta_n=1$ in the right-hand side: the former equality holds because at zero order $T$ does not transform the ghost into itself or into itself plus additional states (for kinematical reasons) and the latter holds because the ghost is the only state with negative norm. Then $|T_{n \alpha}|^2$ is
 the transition probability\footnote{In the present work we define the probability as the absolute value squared of the amplitude. One should keep in mind, however, that other definitions are possible \cite{Salvio:2015gsi}.} for the process $\alpha\rightarrow n$ and the right-hand side of eq. (\ref{optical2})   is  the total probability  that the state $|\alpha\rangle$ decays.  
 
 The next step to compute the ghost decay is to  rewrite the 4-derivative  terms as 2-derivatives where the ghost field is explicit in the Lagrangian. This is the analogue of what we have done in section \ref{The spectrum} to have the field of the scalar $z$ explicit and has been done in Ref. \cite{Hindawi:1995an}. In this case the trick is to introduce a rank two Lagrange multiplier  $\pi_{\mu\nu}$, which is explicitly defined in \cite{Hindawi:1995an}. We eliminate the linear mixing between $\pi_{\mu\nu}$ and the fluctuation of the metric $h_{\mu\nu}$ around the flat space (as well as with all the other fields) by introducing 
 \be \bar{h}_{\mu\nu} \equiv h_{\mu\nu}+\pi_{\mu\nu}+\sqrt{\frac{2}{3}}\frac{\delta z}{\bp} \eta_{\mu\nu},\ee
 where $\delta z\equiv z -\langle z\rangle$. The field $ \bar{h}_{\mu\nu}$ corresponds to the usual massless graviton,  $\pi_{\mu\nu}$ represents the ghost and $\delta z$ the quantum scalar field associated with the fluctuations of $z$ around its vacuum expectation value.  
 
 We also find 
 \be  (\Box -M_2^2)\pi_{\mu\nu} = 0, \qquad  \partial^{\mu} \pi_{\mu\nu}= 0 , \qquad \pi_\mu^{\,\, \mu}= 0 \qquad \quad \mbox{(at the linearized level),}\label{linearized level pi}\ee
 where the indices are raised and lowered here with $\eta_{\mu\nu}$. This confirms that $\pi_{\mu\nu}$ is a massive spin-2 field and can be expanded as follows:
\be \pi_{\mu\nu}(x) = \frac{2}{\bp}\sum_{i, \vec{k}}\frac{1}{\sqrt{2V \omega_\pi(\vec{k})}}\left(e_{\mu\nu}^i(\vec{k})a_{\pi i}(\vec{k}) e^{-ikx}+ \bar e_{\mu\nu}^i(\vec{k})a_{\pi i}^\dagger(\vec{k}) e^{ikx}\right),\label{quantum-pi}\ee
where $V$ is the space volume (that should be taken to $\infty$ at the end), $\omega_\pi(\vec{k})=\sqrt{M_2^2+\vec{k}^2}$ and $e_{\mu\nu}^i(\vec{k})$ are\footnote{$\bar e_{\pi\mu\nu}^i(\vec{k})$ represents the corresponding complex conjugate object.} the polarization tensors corresponding to the wave number $\vec{k}$; the index $i$ labels the helicity state.  The operators $a_{\pi i}(\vec{k})$ and  $a_{\pi i}^\dagger(\vec{k})$ are annihilation and creation operators respectively and fulfill the commutation relations 
\be \left[a_{\pi i}(\vec{k}),a_{\pi j}(\vec{k}')\right]=0,\qquad  \left[a_{\pi i}(\vec{k}),a_{\pi j}(\vec{k}')^\dagger\right] = - \delta_{ij}\delta_{\vec{k}\vec{k}'}.\label{a-comm}\ee 
Note the minus sign on the right-hand side, which has been introduced because we are dealing with ghosts (see Ref. \cite{Salvio:2015gsi}). 

The $e^i_{\mu\nu}(\vec{k})$   transform covariantly as a rank two tensor. Notice that the second equation in  (\ref{linearized level pi}) gives
\be p^\mu e^i_{\mu\nu}(\vec{p}) = 0. \label{gauge-momentum}\ee
In the massive spin-2 case we are considering there are five helicity states (so $i=1,2,3,4,5$). In this case one can easily obtain the $e^i_{\mu\nu}(\vec{p})$ in the rest frame,  $p_\mu=(M_2, 0,0,0)$, and then use general Lorentz boosts to obtain the polarization tensors in an arbitrary frame. In the rest frame  (\ref{gauge-momentum}) becomes $e^i_{0\mu}=0$, which, together with the traceless condition $e^{i\,\, \mu}_{\mu}=0$ (the third equation in  (\ref{linearized level pi})), can be fulfilled by the basis given in Appendix \ref{Polarization tensors}. 
One  can directly check that 
\be e^i_{\mu\nu}(\vec{p})e^{j\mu\nu}(\vec{p}) =\delta^{ij}\ee
and 
\be\sum_{i=1}^5 e^i_{\mu\nu}(\vec{p})\bar e^i_{\rho\sigma}(\vec{p}) = P^{(2)}_{\mu\nu\rho\sigma}(p),\label{sum-helicity} \ee
where 
\be P^{(2)}_{\mu\nu\rho\sigma} = \frac12 T_{\mu\rho}T_{\nu\sigma} + \frac12 T_{\mu\sigma} T_{\nu\rho}-\frac13 T_{\mu\nu}T_{\rho \sigma}, \qquad  \ee
and $T_{\mu\nu}(p) = \eta_{\mu\nu} - p_\mu p_\nu/p^2$. The equality in (\ref{sum-helicity}) can be shown by considering first the rest frame, where the $e^i_{\mu\nu}$ are simple and explicitly given above, and then noticing that both $\sum_{i=1}^5 e^i_{\mu\nu}(\vec{p})\bar e^i_{\rho\sigma}(\vec{p})$ and  $P^{(2)}_{\mu\nu\rho\sigma}(p)$ are rank four tensors and so they coincide if they are equal in a given frame.
 
 Finally, by using the generalized optical theorem in eq. (\ref{optical2}) we obtain that the decay rate $\Gamma$    of the gravitational ghost state with momentum $p_\mu$ is
 \be \Gamma =  \left(\frac{2}{\bp}\right)^2\frac{1}{p_0} \frac15 P^{(2)}_{\mu\nu\rho\sigma}(p) {\rm Im}\, \Pi^{\mu\nu\rho\sigma}(p)\ee
 where $\Pi^{\mu\nu\rho\sigma}$ is the amputated loop Feynman amplitude (multiplied by $-i$). The factor $1/5$ appears because we have averaged with respect to the initial polarization.
 
 That this decay rate is not zero can be explicitly checked by considering   the decay into two real scalars  with mass $m$, for instance two Higgs in the final state. In this case we find  
 \be \Gamma({\rm ghost} \rightarrow {\rm scalar \, \, \, scalar}) = \frac{(M_2^2-4 m^2)^2 {\rm Im} \, B_0(M_2^2, m^2,m^2)}{60 (4\pi)^2 p_0\bp^2}, \ee
where 
the function $B_0$ is defined in Appendix  \ref{appA}, eq. (\ref{B0def}).  In the same  appendix it is also shown
 \be {\rm Im}\, B_0(p^2,m^2,m^2) =\pi\sqrt{1-\frac{4m^2}{p^2}} \theta(p^2-4m^2).\label{ImB0value}\ee
By using this result we find 
 \be \Gamma({\rm ghost} \rightarrow {\rm scalar \, \, \, scalar}) = \frac{(M_2^2-4 m^2)^2\sqrt{1-\frac{4m^2}{p^2}} \theta(M_2^2-4m^2)}{960 \pi p_0\bp^2}. \ee
 This expression can be simplified by going to the ghost rest frame $p=(M_2,0,0,0)$ and assuming $M_2\gg 2m$
  \be \Gamma({\rm ghost} \rightarrow {\rm scalar \, \, \, scalar}) = \frac{M_2^3}{960 \pi \bp^2} \qquad \mbox{(in the ghost rest frame and for $M_2\gg 2m$).}\ee
 
Of course,  the  interactions of $\pi_{\mu\nu}$ with the other fields of the theory under study cannot slower the decay, but they make it faster. This definitely shows that the ghost is unstable and the Lee-Wick idea may be implemented. Other challenges that have to be faced before considering this theory a completely satisfactory UV completion of Einstein's gravity will be mentioned in section \ref{Conclusions and outlook}.

 \subsection{RGEs and threshold effects}

 In order to address naturalness issues in this model (see section \ref{nat}) we need the RGEs: they encode the leading quantum corrections. The one-loop RGEs for the dimensionless couplings for energies above all mass thresholds are  (see e.g. \cite{Avramidi:1985ki, Buttazzo:2013uya,Salvio:2014soa,Salvio:2015cja} and reference therein)
\bea  
\frac{df^2_2}{dt} &=& -\frac{109}{6} f^4_2 , \label{RGEf2}\\
\frac{df^2_0}{dt} &=&  \frac53 \gt^4 + 5 \gt^2 \gs^2 + \frac56 \gs^4 +\gs^4 \frac{(1+6\xi)^2}{3}, \label{RGEf0} \\ 
\frac{d\xi}{dt} &=& (1+6\xi)\left(y_t^2-\frac34 g_2^2 - \frac{3}{20} g_1^2+2\lambda+\frac13\, {\rm Tr}(Y^\dagger Y)\right)\nonumber \\
&&+\frac{\gs^2}{3}\xi(1+6\xi)(2+3\xi) - \frac53 \frac{\gt^4}{\gs^2}\xi, \label{RGExi}  \\
\frac{dg_1^2}{dt}& =&    \frac{41g_1^4}{5}, \qquad  \frac{dg_2^2}{dt} =- \frac{19g_2^4}{3},\qquad \frac{dg_3^2}{dt}= -14 g_3^4, \nonumber \\
\frac{dy_t^2}{dt}   & =& y_t^2\left(9 y_t^2-16g_3^2-\frac{9g_2^2}{2}-\frac{17g_1^2}{10} + 2{\rm Tr}(Y^\dagger Y )+\frac{15}{4} f_2^2\right),\nonumber \\
 \frac{dY}{dt}& =&Y  \left[3 y_t^2-\frac{9}{20} g_1^2-\frac94 g_2^2+\frac32 Y^\dagger Y+ {\rm Tr}(Y^\dagger Y )+\frac{15}{8} f_2^2\right],  \label{RGEY} \\
\frac{d\lambda}{dt}  & =&\left(24\lambda+12y_t^2-\frac{9g_1^2}{5}-9g_2^2+4\, {\rm Tr}(Y^\dagger Y)+5 \gt^2+\gs^2 (1+6\xi)^2\right)\lambda \nonumber\\ &&\hspace{-0.7cm}-\, 6y_t^4 +\frac{9 g_2^4}{8}+\frac{27 g_1^4}{200}+\frac{9 g_2^2 g_1^2}{20} - 2 {\rm Tr}((Y^\dagger Y)^2)+\frac{\xi^2}{2}\left(5 \gt^4+\gs^4(1+6\xi)^2\right), 
\label{eq:RGEnodim0}
\eea
where $t=\ln(\bar{\mu}/\bar{\mu_0})/(4\pi)^2$,  $\bar{\mu}$ is the $\overline{\rm MS}$ renormalization scale\footnote{All renormalized couplings in this work are defined in the $\overline{\rm MS}$ scheme.} and $\bar{\mu}_0$ is a reference energy.  Here we have ignored the Yukawa couplings of the SM that are smaller than the top Yukawa, $y_t$, and the $g_i$ are the gauge couplings.

 Going below the mass thresholds $M_2$ and $M_0'$ one can neglect the contributions due to $f_2$ and $f_0$ respectively. One can wonder whether the scalar threshold due to $z$ induces a tree-level shift of the quartic Higgs coupling, along the lines of \cite{RandjbarDaemi:2006gf}. We now show that such a shift is negligible. This effect can emerge because setting the heavy scalar (in this case $\delta z \equiv z - \langle z\rangle$) equal to zero is not compatible with the equations of motion. This occurs if there are scalar couplings of the schematic form (heavy) x (light) x (light) in the Lagrangian (in this case $\delta z \delta h^2$, where $\delta h \equiv h - \sqrt{2}v$). Using Eq. (\ref{scalar-EF}) leads to such a coupling, $\sim f_0^2 \xi \bp  \,  \delta z \delta h^2$ (modulo order one factors and neglecting contributions suppressed by $v^2/\bp^2$ and  the tiny value of $\Lambda$). 
The shift $\delta \lambda$ in the quartic coupling is given by the square of the coefficient of the (heavy) x (light) x (light) term, in this case $\sim f_0^2 \xi \bp$ times the propagator of $z$ at zero external momentum \cite{RandjbarDaemi:2006gf}: $ \delta \lambda \sim f_0^2 \xi^2$. 
In sections \ref{nat} and \ref{inf} we will see that the requirement of successful inflation and Higgs mass naturalness implies $f_0 \sim 10^{-5}$ and $\xi \approx -1/6$, so that $f_0^2 \xi^2 \sim 10^{-11}$ and this effect is negligibly small.

As far as the RGEs of  the mass parameters are concerned, we find that $m^2$,  $M$ and $\bp^2$ obey
\bea \frac{dm^2}{dt}&=& m^2 \left(12 \lambda +6 y_t^2 -\frac92 g_2^2 -\frac{9}{10} g_1^2+2 {\rm Tr} \left(Y^\dagger Y\right)
   +5 f_2^2+ f_0^2 (1+6\xi)+6\xi^2+2G\right) \nonumber \\\ && + 8 {\rm Tr}\left(Y^\dagger Y M^\dagger M\right)  -5f_2^4 \xi \bp^2-f_0^4\xi(1+6\xi)\bp^2,  \label{RGEm2} \\ \frac{dM}{dt} &=&  M   Y^\dagger Y+   \left(Y^{\dagger} Y\right)^T M 	
 + \frac{15}{8}f_2^2 M+MG,\label{RGEM} \\ 
    \frac{d \bp^2 }{dt} & =& -\frac{2}{3} m^2+\frac{1}{24}{\rm Tr}(M^\dagger M) -4\xi m^2  +\left(\frac{2f_0^2}{3}-\frac{5f_2^4}{3f_0^4}+2G\right)\bp^2.\label{RGEPlanck}
    \eea 
Here $G$ is a gauge-dependent quantity: for example, using the same gauge fixing action as in Ref. \cite{Salvio:2014soa},
\be \label{eq:gf}
S_{\rm gf} = - \frac{1}{2\xi_g}\int d^4x ~ f_\mu \partial^2 f_\mu,\qquad
f_\mu = \partial_\nu( h_{\mu\nu} - c_g \frac12 \eta_{\mu\nu} h_{\alpha \alpha}),
\ee
where $h_{\mu\nu} = g_{\mu\nu}-\eta_{\mu\nu}$, leads to 
\be G= \frac{(3c_g^2-12c_g+13)\xi_g}{4(c_g-2)}+\frac{3(c_g-1)^2f_0^2}{4(c_g-2)^2}.\ee
The gauge dependence cancels as it should in the RGEs of $M/\bp$ and $m^2/\bp^2$.

\section{Naturalness} \label{nat}

Notice that the $\beta$-function in (\ref{RGEM}) vanishes as $M\rightarrow 0$. Therefore, by starting from small values (such as those in (\ref{RHvalue})) at low energy, one does not end up with a much larger $M$. This occurs because $M$ breaks lepton symmetry, while all other fields (gravity included) preserves it. Such small values of $M$ are therefore natural even taking into account gravity. This is because our softened gravity preserves  global lepton symmetry.

The same is true for $Y$, given the structure of the $\beta$-function in (\ref{RGEY}). As a result, the naturalness of the EW scale $m$ leads to the same conditions obtained in   \cite{Salvio:2014soa} (as it can be seen from Eq. (\ref{RGEm2})): the order of magnitude of $f_2^4$ and $f_0^4(1+6\xi)$ should not exceed $M_h^2/\bp^2$. This  condition  is preserved by the RG-running (see Eqs. (\ref{RGEf2}), (\ref{RGEf0})  and  (\ref{RGExi})). The smallness of these couplings corresponds to the softening  of gravity. Notice that one important ingredient to ensure this result is the fact that the small values of $M$ and $Y$ (see for example (\ref{RHvalue})) ensures that neutrinos do not give unnaturally large corrections to the Higgs mass. This is opposed to the standard leptogenesis scenario \cite{Fukugita:1986hr}, which occurs through the decay of very heavy \cite{Davidson:2002qv} right-handed neutrinos  and can introduce a fine-tuning in the Higgs mass \cite{Farina:2013mla}. 

In a similar way we also show now that the smallness of the $\theta$-term is natural in this context.  In the SM the $\beta$-function of $\theta$ starts at least at 7 loops and is at most of order $10^{-15}$ \cite{Burgess:2007zi}. This is because one needs to construct a flavor invariant CP-breaking term out of the quark Yukawa couplings. Therefore in the SM the running of $\theta$ is negligibly small. The right-handed neutrino sector contains other sources of CP breaking and can potentially introduce a larger running.  However, in order to connect the $\theta$-vertex with a right-handed neutrino you need three loops (you should insert a quark,  a Higgs and  a right-handed neutrino). This leads to a $1/(4\pi)^6$ suppression. Moreover, you have at least an extra factor that is quadratic in the $Y_{ij}$, which have to be very small given that all right-handed neutrinos are below the EW scale (see e.g. Eq. (\ref{RHvalue})). Therefore, also the right-handed neutrino sector preserves the smallness of the $\theta$-running. Finally, notice that gravity, given that it is softened and CP-preserving in our context, does not reintroduce a sizable running.

It is now a good point to comment on the possible CP-breaking extension of the softened gravity theory at hand. Given that we limit to terms in the Lagrangian which are at most quadratic in the curvature, we could add a ``gravitational $\theta$-term":
\be \epsilon^{\mu\nu\rho\sigma}R_{\mu\nu}^{\,\, \,\,\,\, \alpha\beta}R_{\rho\sigma \alpha\beta}, \label{gravTheta}\ee
which may potentially affect the observable predictions of the theory. This term (as well as some phases in the quark mass matrix) could be (partially) due to removing the QCD $\theta$-term via a chiral transformation of the quarks \cite{Delbourgo:1972xb}. Since it can be rotated away with an anomalous chiral transformation its coefficient in the Lagrangian cannot be much larger than $2\pi$ times $1/(4\pi)^2$. For such a small value the gravitational $\theta$-term could only affect very energetic phenomena, such as inflation.  Moreover, the presence of right-handed neutrinos helps: one can perform a U(1) transformation of $N_i$, that is $N_j\rightarrow \exp(i \beta) N_j$, where $\beta$ is a real number, which removes completely such gravitational term and, as side effect, only rescales $M_j$ and $Y_{ij}$ in the following way
$M_j\rightarrow \exp(2 i \beta)  M_j,$ $Y_{ij} \rightarrow  \exp(i \beta)  Y_{ij}.$
This rescaling produces no effect in neutrino observables, which can therefore be compatible with data. This is because the neutrino mass matrix in (\ref{see-saw})    is invariant under such transformation.

\section{Inflation}\label{inf}

Let us finally turn to inflation. In a similar context \cite{Kannike:2015apa,Salvio:2015kka,Salvio:2015jgu} showed that inflation is mainly triggered by the Starobinsky effective scalar $z$, rather than the Higgs. We show that the same happens here even in the natural parameter space. 

\subsection{Multifield inflation formalism}\label{Multifield inflation formalism}

Although the theory studied here  has only two scalar fields (in the unitary gauge), it is convenient to start from a formalism suitable for a general number of scalars (as the equations will be shorter). This formalism has been studied in the past \cite{info}, but we will obtain  weaker slow-roll conditions, therefore we present the explicit derivation here. We also take advantage to elucidate some points in multifield inflation.

We rewrite the scalar-tensor Lagrangian in (\ref{eq:Einstein}) as 
\be 
\Lag_{\rm st} =\sqrt{- g_E} \bigg[  \frac{\frac13 R_E^2 -  R_{E\mu\nu}^2}{f_2^2}
 - \frac{\bar M_{\rm Pl}'^2}{2} R_E + 
\frac{K_{ij}(\phi) }{2}\partial_\mu \phi^i\partial^\mu \phi^j-
U(\phi)  \bigg],
\ee
where  the field metric $K_{ij}$ and the potential $U$ are generic functions of the scalar fields $\phi^i$. In our case we are interested in the case $\phi^i = \{z,h\}$, where $h$ is the physical Higgs field, but we keep the formalism general as explained above. 

Next we consider the Friedmann-Robertson-Walker (FRW) metric.
 \be ds^2_E\equiv g^E_{\mu\nu}dx^\mu dx^\nu = dt^2 -a(t)^2 \left[\frac{dr^2}{1- k r^2} +r^2(d\Theta^2 +\sin^2\Theta \, d\phi^2)\right].  \ee
We neglect from now on the curvature contribution $k$ as during inflation the energy density is dominated by the scalar fields. Then the Einstein equations (EEs) and the scalar equations imply the following equations for $a(t)$ and the spatially homogeneous fields $\phi^i(t)$  
  \bea H^2&=&\frac{K_{ij} \dot\phi^i\dot \phi^j/2+U}{3 \bar M_{\rm Pl}'^2} , \qquad \mbox{($tt$-component of EEs) } \label{EHt}\\ 
  2 a(t) \ddot a(t)+\dot a(t)^2
  &=& 
  \frac{a^2(K_{ij} \dot\phi^i\dot \phi^j/2-U)}{\bar M_{\rm Pl}'^2(k r^2-1)} , \qquad \mbox{($rr$-component of EEs) }
  \\ 
   2 a(t) \ddot a(t)+\dot a(t)^2 &=&\frac{a^2 (U-K_{ij} \dot\phi^i\dot \phi^j/2)}{\bar M_{\rm Pl}'^2} , \qquad \mbox{($\theta \theta$-component of EEs) }\\ 2 a(t) \ddot a(t)+\dot a(t)^2 &=&\frac{a^2 (U-K_{ij} \dot\phi^i\dot \phi^j/2)}{\bar M_{\rm Pl}'^2} , \qquad \mbox{($\phi \phi$-component of EEs) }\\ \ddot \phi^i +\gamma^i_{jk}\dot \phi^j\dot \phi^k +3H\dot \phi^i+U^{,i}  &=&0,\label{scalarEq} \qquad \mbox{(scalar field equation) }\eea
 where   $H\equiv \dot a/a$ and a dot denotes a derivative with respect to $t$. Also for a generic function $F$ of the scalar fields, we defined $F_{,i}\equiv \partial F/\partial \phi^i$,   the affine connection $\gamma^i_{jk}$  in the scalar field space is
 \be \gamma^i_{jk}\equiv \frac{K^{il}}{2}\left(K_{lj,k}+K_{lk,j}-K_{jk,l}\right) \ee
 and $K^{ij}$ denotes the inverse of the field metric (which is used to raise and lower the scalar indices $i,j,k, ...$); for example $F^{,i}\equiv K^{ij}F_{,j}$. The $rr$-, $\theta \theta$- and $\phi \phi$-components of the EEs are only one independent equation, thus the EEs can be simplified to 
  \bea H^2&=&\frac{K_{ij} \dot\phi^i\dot \phi^j/2+U}{3 \bar M_{\rm Pl}^2} , \\ 
 \dot H&=& - \frac{K_{ij} \dot \phi^i\dot \phi^j}{ 2\bar M_{\rm Pl}^2}. \eea
 
 Notice that  the term suppressed by $f_2^2$ in the Lagrangian has no effect because it is equal to (modulo total derivatives) to the square of the  Weyl  tensor which vanishes on the FRW metric. We assume that term has no effect on inflation at the quantum level either\footnote{This is the case when $M_2$ is roughly above the Hubble rate during inflation $H_{\rm inf}$, otherwise there are two modifications in the analysis below\cite{Salvio:2017xul} (see also \cite{Clunan:2009er,Deruelle:2012xv,Ivanov:2016hcm} for previous less general results):
 \begin{itemize}
 \item there is an extra iscurvature scalar mode corresponding to the helicity-0 component of the spin-2 massive ghost (which, however, satisfies the most recent bounds on isocurvature power spectra \cite{Planck2015});
 \item  one has to take into account an extra suppression factor $1/(1+2H_{\rm inf}^2/M_2^2)$ in front of the tensor-to-scalar ratio $r$, which will be given in eq. (\ref{r}). Given that the observations only give us an upper bound on $r$, this modification leaves the model viable.
 \end{itemize}}. 

\subsubsection{The slow-roll approximation}
We now describe the slow-roll approximation within this formalism. The scalar fields roll slowly down the potential when 
\be K_{ij} \dot\phi^i\dot \phi^j \ll U, \quad  |\ddot \phi^i +\gamma^i_{jk}\dot \phi^j\dot \phi^k|\ll  3H \dot \phi^i,\quad |\ddot \phi^i +\gamma^i_{jk}\dot \phi^j\dot \phi^k|\ll   U^{,i}.\label{slow-roll-cond}\ee
Then from Eqs. (\ref{EHt}) and (\ref{scalarEq}) we obtain
\be H^2 \simeq \frac{U}{3  \bar M_{\rm Pl}^2} ,\qquad  \dot \phi^i\simeq -\frac{U^{,i}}{3H}. \label{slow-roll-eq}\ee
(in our notation $U^{,i}\equiv K^{ij} U_{,j}$, where $U_{,i}\equiv \partial U/\partial \phi^i$).
By using  (\ref{slow-roll-eq}) in the first condition in (\ref{slow-roll-cond}) we obtain  {\it the first slow-roll condition}
\be \epsilon \equiv  \frac{  \bar M_{\rm Pl}^2 U_{,i}U^{,i}}{2U^2} \ll  1. \label{1st-slow-roll}\ee 
Eq.  (\ref{slow-roll-eq})  tell us 
\be  \frac{\dot H}{H^2}\simeq -\epsilon, \label{eqHepsilon}\ee
which is guaranteed to be small by (\ref{1st-slow-roll}).
From   (\ref{slow-roll-eq})  we find
\be \frac{\ddot \phi^i +\gamma^i_{jk}\dot \phi^j\dot \phi^k}{H\dot \phi^i}\simeq - \frac{\dot H}{H^2}- \frac{\bar M_{\rm Pl}^2U^{;i}_{\, \, \, ;j} U^{,j}}{U U^{,i}}\simeq - \frac{\bar M_{\rm Pl}^2U^{;i}_{\, \, \, ;j} U^{,j}}{U U^{,i}}, \qquad \mbox{($i$  not summed) }\ee
where for a generic vector $V^i$ on the scalar field space, we defined the covariant derivative $V^i_{\, \, \,;j}\equiv \partial V^i/\partial \phi^j+\gamma^i_{jk} V^k$. Notice that in the formula above the index $i$ is not summed and in the last step we have neglected $\dot H/H^2$ that  we have just proved to be small. Therefore, from (\ref{slow-roll-cond}) we obtain {\it the second slow-roll condition }
\be \left|\frac{\eta^{i}_{\,\,\, j} U^{,j}}{U^{,i}}\right|  \ll  1 \quad \mbox{($i$  not summed), }\quad \mbox{where}\quad \eta^{i}_{\,\,\, j}\equiv \frac{\bar M_{\rm Pl}^2 U^{;i}_{\,\,\, ;j}}{U}  \label{SlowRoll2} \ee
 It is easy to check that $\epsilon$ and $\eta^i_{\,\,\, j}$ reduce to the well-known single field slow-roll parameters in the presence of only one field. The second slow-roll condition  is weaker than the one found in \cite{info} where it is assumed $|\eta^{i}_{\,\,\, j}| < 1$ and $U^{,j}/U^{,i}$ of order one.

Combining the two equations in (\ref{slow-roll-eq}) we obtain the following dynamical system for $\phi^i$: 
\be \dot \phi^i=-\frac{\bar M_{\rm Pl}U^{,i}(\phi)}{\sqrt{3U(\phi)}}, \label{dynamical1}\ee
which can be solved with a condition at some initial time  $t_0$: that is $\phi^i(t_0)=\phi^i_0$. Once the functions $\phi^i(t)$ are known we can obtain $H(t)$ from the first equation in (\ref{slow-roll-eq}).
Let us introduce the number of e-folds $N$ by 
\be N(\phi_0) \equiv \int_{t_e}^{t_0(\phi_0)} dt' H(t'), \label{Ndef}\ee
where $t_e$ is the time when inflation ends. Dropping the label on $t_0$ and $\phi_0$ as they are generic values we have
\be N(\phi) \equiv \int_{t_e}^{t(\phi)} dt' H(t'). \label{Ndef2}\ee
Notice that we write that $t$ is a function of $\phi$: this is because once the initial position $\phi$ in field space is fixed the time required to go from $\phi$ to the field value when inflation ends is fixed too because the dynamical system in (\ref{dynamical1}) is of the first order. Note, however, that $H$ also generically depends on $\phi$.
 Definition (\ref{Ndef2}) implies
\be \frac{dN}{dt}=H,\label{dN/dt}\ee which can be used in (\ref{slow-roll-eq}) to obtain a simpler dynamical system for $\phi^i$ where the independent variable is $N$ instead of $t$: 
\be \frac{d\phi^i}{dN}=- \frac{\bar M_{\rm Pl}^2U^{,i}(\phi)}{U(\phi)}. \label{SlowRollEqN}\ee 

\subsubsection{Observable predictions}

One can then extract predictions for observable quantities such as the power spectrum  $P_\mathcal{R}(k)$ of scalar fluctuations, the spectral index $n_s$ and the tensor-to-scalar ratio $r$. The measured values at $k= 0.002\, \mbox{Mpc}^{-1}$ are
$ P_\mathcal{R}(k) = (2.14 \pm  0.05) \times 10^{-9}$~\cite{Planck2015},
$n_s=   0.965 \pm 0.006$~\cite{Ade:2013uln,Planck2015} and $r< 0.09$  \cite{Planck2015}. The power spectrum $P_\mathcal{R}(k)$ is (see eq. (40) of \cite{Sasaki:1995aw})
\be P_\mathcal{R}(k)=\left(\frac{H}{2\pi}\right)^2 N_{,i}N^{,i},\label{power-spectrum}\ee
computed at horizon exit $k=aH$.   The spectral index $n_s$ of scalar perturbations can be computed as 
\be n_s=1+\frac{d\ln P_\mathcal{R}}{d\ln k}\ee
By using now $d\ln k=d\ln aH\simeq H dt$, where we noticed that during a nearly exponential expansion $\dot a/a\simeq \ddot a/\dot a$, and eq. (\ref{power-spectrum}) we find
\be n_s\simeq 1 +2\frac{\dot H}{H^2} +2\frac{N^{,i}\dot N_{,i}}{H N_{,j}N^{,j}} \, . \label{nsFormula}\ee
where $N$ is the quantity defined in (\ref{Ndef2}). The second term on the right-hand side can be substituted by $-2\epsilon$ (eq. (\ref{eqHepsilon})), while the third one can be computed by using
\bea \dot N_{,i}&=&  \dot N_{;i}=\dot \phi^{j} N_{;i,j}= (\dot \phi^j N_{,j} )_{;i} -(\dot \phi^j)_{;i} N_{,j}= H_{:i} -N^{,j} (\dot \phi_j)_{;i}\, , \\
(\dot \phi_j)_{;i}&\simeq &-\left(\frac{U_{,j}}{3H}\right)_{;i} = \frac{H_{,i}U_{,j}}{3H^2}-\frac{U_{;j;i}}{3H}.\label{phiji}\eea 
This leads to
\be   n_s =1-2\epsilon - \frac{ 2 }{ \bar M_{\rm Pl}'^2 N_{,i}N^{,i}}+\frac{2\eta_{ij}N^{,i}N^{,j}}{N_{,k}N^{,k}} 
\ee
where we used (\ref{slow-roll-eq}). This formula does not contain a term with the Riemann tensor, unlike eq. (42) of \cite{Sasaki:1995aw} 
\be n_s =1+\frac{2\dot H}{H^2} - \frac{ 2 }{ \bar M_{\rm Pl}'^2 N_{,i}N^{,i}}+\frac{2\eta_{ij}N^{,i}N^{,j}}{N_{,k}N^{,k}} -\frac{2 \bar M_{\rm Pl}'^4R_{ijkl}N^{,i}N^{,l} U^{,j}U^{,k}}{3 N_{,m}N^{,m} U^2}, 
 \ee 
 because the slow-roll eqs. (\ref{slow-roll-eq}) have been used to evaluate  $(\dot \phi_j)_{;i}$ in eq. (\ref{phiji}).

Also the tensor-to-scalar ratio can now be easily computed by using a textbook formula for the tensor power spectrum 
\be P_t(k) = \frac{2}{\bar M_{\rm Pl}^2} \left(\frac{H}{2\pi}\right)^2\ee 
to obtain
\be r\equiv \frac{4 P_t(k)}{P_\mathcal{R}(k)}=\frac{8}{ \bar M_{\rm Pl}^2 N_{,i}N^{,i}}. \label{r}\ee

\subsection{Higgs-Starobinsky system}\label{Higgs-Starobinsky system}

\begin{figure}[t]
\begin{center}\includegraphics[scale=0.71]{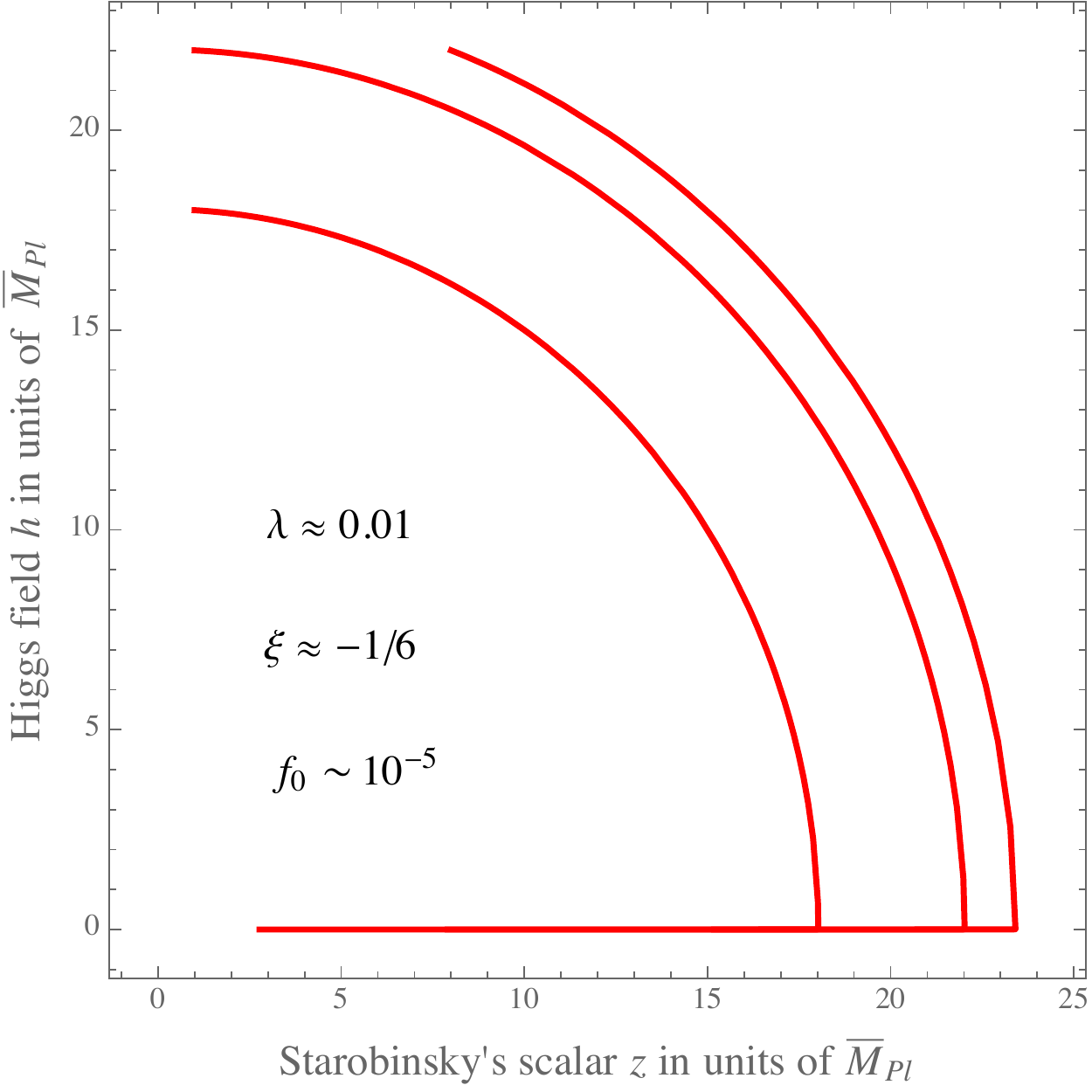} \end{center}
\vspace{-0.4cm}
   \caption{{\small \it Inflationary path of the scalar fields. } }
\label{attractor}
\end{figure}
We now apply the formalism of section \ref{Multifield inflation formalism} to the Higgs-Starobinsky system defined by (\ref{eq:Einstein}) and (\ref{scalar-EF}).

A qualitative analysis of the potential $U=V_E$ shows that inflation is mainly triggered by $z$ because, even if  $\lambda > 0$, namely the case in which Higgs inflation \cite{Bezrukov:2007ep} is possible \cite{Bezrukov:2008ej,Bezrukov:2009db,Salvio:2013rja},  the fields $\{z, h\}$ are rapidly attracted to the line $h=0$ (the running forces $\lambda$ to sizable values and the potential barrier for $h\neq 0$ is too steep to allow Higgs inflation). Matching the observed $P_\mathcal{R}(k)$, leads to $f_0 \sim 10^{-5}$ and therefore $\xi\approx -1/6$ to achieve Higgs mass  naturalness, as discussed in section \ref{nat}. Note that inserting $f_0 \sim 10^{-5}$ in $M_0'$ leads to $M_0' \gg m$ as we anticipated in section \ref{The spectrum}.

 In figure \ref{attractor} we show the presence of the above-mentioned attractor even for the natural values    $\xi\approx -1/6$ and $f_0 \sim  10^{-5}$. It is only when the fields reach the attractor that the slow-roll conditions in (\ref{1st-slow-roll}) and (\ref{SlowRoll2}) are satisfied.  The curves have been obtained by solving the field equations (\ref{SlowRollEqN}) in the slow-roll approximation. 
 
By using Eqs. (\ref{nsFormula}) and (\ref{r}) we obtain the predictions for the observable quantities, $n_s \approx 0.97$ and $r\approx 0.0031$ for a number of e-folds $N\approx 59$. We obtain results that are very close to Starobinsky's predictions $n_s \approx 0.97$ and $r\approx 0.0035$ \cite{Kannike:2015apa} for the reasons explained above.  It is interesting to note that the softened gravity theory under study gives a justification for the $R^2$ term in the Lagrangian: as we have previously mentioned, even if we do not introduce it in the classical Lagraingian, quantum corrections generate it.

  For negative values of $\lambda$ at the inflationary scales, which are suggested by recent calculations \cite{Bezrukov:2012sa, Degrassi:2012ry, Buttazzo:2013uya},  one should instead require directly that the Higgs is at the EW minimum of the SM potential\footnote{The quantum and thermal probability of jumping to the true minimum has been recently considered even in the presence of the extra gravitational terms in Eq. (\ref{gravity-Lag}) \cite{Salvio:2016mvj}. The conclusion is that the lifetime of the EW vacuum can  well be much bigger than the age of the universe without tension with any experiment.}: large field values of the Higgs above the SM potential barrier would lead to a run away for the Higgs field, which would not eventually roll towards the EW vacuum. 
  
   Therefore the inflationary nature of the model is close to that of Starobinsky's inflation, in good agreement with current cosmological observations: the differences with Starobinsky's predictions are within current uncertainties, but future observations may give us more information (we will discuss this point in section \ref{Conclusions and outlook})

\section{Conclusions and outlook}\label{Conclusions and outlook}

We have presented a softened gravity theory that, besides having a natural Higgs mass, also possesses a natural QCD $\theta$ angle. Three right-handed neutrinos below the EW scale can explain the neutrino oscillations, DM and BAU and, at the same time, protects the theory from gravitational violation of CP invariance.  Contrary to the standard lore, we have shown (within the softened gravity theory studied here) that the smallness of the right-handed Majorana neutrino masses and Yukawa couplings required to explain neutrino oscillations, DM and BAU is not a fine-tuning: their smallness is preserved by the RGEs (which we have determined explicitly), even if the gravity corrections are included. Moreover, low-scale right-handed neutrinos give negligible corrections to the Higgs mass. Therefore,  the softened gravity idea is not separate from the idea of low-scale neutrinos, these two ingredients mutually reinforce each other.

The implementation of  softened gravity that we have used here present a spin-2 heavy ghost in its spectrum. We have shown that this is an unstable state and therefore the basic condition for a Lee-Wick interpretation of this theory is fulfilled. Open problems for the future include the non-perturbative formulation of the theory and the study of its causal structure (which could perhaps be done following the ideas of \cite{Salvio:2015gsi} and \cite{Grinstein:2008bg}, respectively).

Moreover, we have shown that inflation and its predictions are close to those of Starobinsy's $R^2$ model. Given the current observational uncertainties they look   actually the same. However, we have found some differences such that they can be distinguished by future observations such as CMBpol \cite{Baumann:2008aq}. For instance, future sensitivity for $r$ can well be at the level of $10^{-3}$ or below \cite{Baumann:2008aq,Creminelli:2015oda}. We hope that the present work can stimulate future experimental as well as theoretical efforts in distinguishing these theories. A future theoretical goal could be for instance an improved calculation of the inflationary predictions by going beyond the slow-roll approximation. 

\vspace{0.3cm}
\noindent {\bf Acknowledgments.} We thank A. Strumia, G. Villadoro and M. Shaposhnikov for useful discussions. This work was supported by the grant 669668 -- NEO-NAT -- ERC-AdG-2014.
 
   \appendix 
   
   \section{Polarization tensors}\label{Polarization tensors}
   
 The polarization tensors for a massive spin-2 field are  \cite{Veltman}
\be \label{basis} \bac \mbox{(in the rest frame)}\qquad e^1_{\mu\nu}=\frac{1}{\sqrt2} \left(\bacccc 0 & 0 & 0 & 0\\
0 & 1 & 0 & 0\\ 0 & 0 & -1 & 0\\ 0 & 0 & 0 & 0 \ea \right), \quad e^2_{\mu\nu}=\frac{1}{\sqrt2} \left(\bacccc 0 & 0 & 0 & 0\\
0 & 0 & 1 & 0\\ 0 & 1 & 0 & 0\\ 0 & 0 & 0 & 0 \ea \right),
\\
e^3_{\mu\nu}=\frac{1}{\sqrt2} \left(\bacccc 0 & 0 & 0 & 0\\
0 & 0 & 0 & 1\\ 0 & 0 & 0 & 0\\ 0 & 1 & 0 & 0 \ea \right), \quad e^4_{\mu\nu}=\frac{1}{\sqrt2} \left(\bacccc 0 & 0 & 0 & 0\\
0 & 0 & 0 & 0\\ 0 & 0 & 0 & 1\\ 0 & 0 & 1 & 0 \ea \right), \quad e^5_{\mu\nu}=\frac{\sqrt2}{\sqrt3} \left(\bacccc 0 & 0 & 0 & 0\\
0 & 1/2 & 0 & 0\\ 0 & 0 & 1/2 & 0\\ 0 & 0 & 0 & -1 \ea \right).\nonumber \ea \ee

 \section{Imaginary part of loop functions} \label{appA}


  By using the   formula 
\be \frac1{ab} = \int_0^1 \frac{dx}{\left[b+(a-b) x\right]^2} \ee
we find that the function $B_0$, which is defined
by 
\be B_0(p^2,m_1^2,m_2^2)\equiv \frac{1}{i\pi^2}\int d^dq\frac{1}{(q^2-m_1^2+i\epsilon)\left[(q+p)^2-m_2^2+i\epsilon\right]}\label{B0def}\ee
 in dimensional regularization (with space-time dimension $d$), can be written as 
\be B_0(p^2,m_1^2,m_2^2)= \frac{1}{i\pi^2}\int d^dq\int_0^1 dx \frac{1}{\left[q^2-m_1^2+\left(p^2+2 q p +m_1^2-m_2^2\right)x+i\epsilon\right]^2}\ee
and then by  introducing the new loop variable $k=q+x p$
\be B_0(p^2,m_1^2,m_2^2)\equiv \frac{1}{i\pi^2}\int d^dk\int_0^1 dx \frac{1}{\left[k^2+x^2 p^2-m_1^2+\left(p^2-2  x p^2  +m_1^2-m_2^2\right)x+i\epsilon\right]^2}\ee
and setting $m_1=m_2	\equiv m$
\be B_0(p^2,m^2,m^2)\equiv \frac{1}{i\pi^2}\int d^dk\int_0^1 dx \frac{1}{\left[k^2-F(p^2,x)+i\epsilon\right]^2},\ee
where
\be F(p^2,x)\equiv m^2+ x(x-1)p^2.\ee
${\rm Im}\, B_0(p^2,m^2,m^2)\neq 0$ only when $F(p^2,x)$ is negative, that is for $x_-< x< x_+$ where $x_\pm=(1\pm\sqrt{1-4m^2/p^2})/2$, which is possible only for $p^2>4m^2$. By performing the Wick rotation one therefore finds 
\be {\rm Im}\, B_0(p^2,m^2,m^2)=\frac1{\pi^2} \int_{x_-}^{x_+} dx\,\, {\rm Im} \int d^4 k_E \frac{\theta(p^2-4m^2)}{(-k_E^2+|F|+i\epsilon)^2},
\ee
where the label $E$ reminds us that we are now in the Euclidean space (not to be confused with the Einstein frame label). By using spherical coordinates in this space we obtain
\be {\rm Im}\, B_0(p^2,m^2,m^2)=2 \theta(p^2-4m^2) \int_{x_-}^{x_+} dx \,\, {\rm Im} \int_0^\infty \frac{dr r^3}{(-r^2+|F|+i\epsilon)^2} . \label{ImB01}\ee
We have 
\be  \int  \frac{dr r^3}{(-r^2+|F|+i\epsilon)^2} = \frac12 \left(\frac{|F|+i\epsilon}{|F|+i\epsilon-r^2}+\log\left(|F|+i\epsilon-r^2\right)\right).\ee
We can now split the integral $\int_0^\infty dr$ in the integral $\int_0^{\sqrt{|F|+\delta}} dr$ plus $\int^\infty_{\sqrt{|F|+\delta}} dr$, where $\delta$ is a positive number, and notice that only the former can give rise to an imaginary part. So 
\be  {\rm Im} \int_0^\infty \frac{dr r^3}{(-r^2+|F|+i\epsilon)^2} =\frac12 {\rm Im} \log(-\delta+i\epsilon) =\frac12 {\rm Im} \log(-1)= \frac\pi{2}\ee
By inserting this result in (\ref{ImB01})  we find 
\be {\rm Im}\, B_0(p^2,m^2,m^2)= \pi \sqrt{1-\frac{4m^2}{p^2}} \theta(p^2-4m^2). \ee

\vspace{0.3cm}

 \begin{multicols}{2}

\end{multicols}

\end{document}